\pdfoutput=1
\documentclass[10pt]{article}

\usepackage[T1]{fontenc}
\usepackage{lmodern}
\usepackage[a4paper,total={170mm,257mm},left=20mm,top=20mm]{geometry}
\usepackage{authblk}
\usepackage{amsmath,amssymb}
\usepackage{graphicx}
\usepackage{xspace}
\usepackage[nocompress]{cite}
\usepackage{xurl}
\usepackage{microtype}
\usepackage[hidelinks]{hyperref}

\newcommand{\model}{\textnormal{\textsc{QSoftCore}}\xspace}
\newcommand{\qcs}{\texttt{QCS}\xspace}
\let\originaltexttt\texttt
\renewcommand{\texttt}[1]{{\footnotesize\originaltexttt{#1}}}
\hypersetup{
  pdftitle={QSoftCore: A Conceptual Foundation for Quantum Software Engineering},
  pdfauthor={Tao Yue, Yuechen Li, and Man Zhang}
}

\title{\model: A Conceptual Foundation for Quantum Software Engineering}
\author[1]{Tao Yue}
\author[1]{Yuechen Li}
\author[1]{Man Zhang}
\affil[1]{Beihang University, China}
\affil[ ]{\texttt{yuetao@buaa.edu.cn; liyuechen@buaa.edu.cn; manzhang@buaa.edu.cn}}
\date{}

\begin{document}
\maketitle

\begin{abstract}
Quantum software engineering (QSE) requires understanding application requirements, quantum programs and algorithms, quantum computation paradigms, executions, and deployment. However, relevant quantum computing (QC) concepts are often defined across physical, mathematical, and paradigm-specific levels of abstraction, making it difficult for the QSE community and hence proposed methods to refer to them consistently. To this end, we present \model, a conceptual model that reorganizes established QC concepts from a software engineering perspective and connects quantum applications, computational problems, algorithms, configurations, programs, and paradigm-specific computation specifications with state evolution, execution results, and backends. Our assessment against QSE literature across key software engineering activities indicates that \model provides shared QC referents for quantum computations, executions, deployment configurations, backends, and measurement evidence across the QSE lifecycle.
\end{abstract}


\section{Introduction}
Quantum computing (QC) is becoming a software-intensive discipline: applications increasingly rely on programming languages, compilers, software development kits, workflow engines, cloud platforms, simulators, quantum processing units (QPUs), and classical components. Accordingly, quantum software engineering (QSE) now addresses activities across the software lifecycle~\cite{piattini2022landscape,weder2022qsdl,dwivedi2024qsesurvey,murillo2025qseroadmap}, including requirements~\cite{yue2023towardsquant, stein2024requirementsmanufacturing}, architecture~\cite{ahmad2024qcaas,aktar2025archdecisions}, implementation, testing~\cite{wang2024testcasemini,li2026dynamicoracle}, and debugging~\cite{li2020proq, zhao2023bugs4qabench}.

These activities must relate application requirements to a computational problem and an algorithmic choice, distinguish a source program from the computation it constructs, connect a logical specification to its compiled and deployed realization, and interpret execution results under sampling, noise, calibration, and backend assumptions. Existing QC explanations, however, are commonly structured around subjects weakly aligned with classical software engineering, such as physical and mathematical conceptions, abstruse algorithmic principles, or particular computation paradigms~\cite{nielsen2010quantum}. QSE studies consequently tend to introduce the QC concepts needed for an individual task, often with a gate-based and circuit-centered focus~\cite{garcia2023qst,fortunato2024vv,murillo2025qseroadmap}. 

To address this gap, this paper presents \model, a conceptual model that reorganizes established QC concepts from a software engineering (SE) perspective. Rather than introducing a new quantum formalism or reproducing general SE taxonomies, \model organizes QSE-relevant QC concepts into five views: 1) applications and software systems, linking requirements and computational problems to algorithms, configurations, and software components; 2) computation specifications for gate-based, analog, quantum-annealing, and measurement-based paradigms; 3) gate-based computation, detailing the circuits, operations, qubits, and states on which most current QSE work rests; 4) execution and evolution, separating the intended computation from a concrete run and its runtime evidence; and 5) deployment, relating logical specifications to backend-specific artifacts, including mappings, calibration, noise, and physical resources. Two worked application examples then connect these views.

We assess \model against the QSE literature: across the activities the literature addresses, we check whether the QC concepts required are represented by \model, distinguishing QC concepts (e.g., a quantum circuit) from task-specific SE artifacts (e.g., a test case). Across requirements, architecture design, code generation, execution and deployment, and quality assurance, the assessment finds that the QSE literature is predominantly gate-based and circuit-centered, remains close to programming-level abstractions, and offers limited support for application-level and hybrid design, alternative computation paradigms, deployment reasoning, and interpreting simulator- and QPU-derived evidence. \model therefore covers the QC concepts these works use and adds explicit representations for alternative paradigms, quantum evolution, deployment context, execution evidence, and raw versus mitigated results, offering QSE methods and tools a shared QC vocabulary for defining and tracing artifacts across activities.
By giving QSE a shared QC vocabulary, \model allows researchers, tool builders, and practitioners to describe algorithms, programs, executions, and deployment contexts in comparable terms, so that artifacts, assumptions, and evidence can be reused across methods, tools, and studies rather than re-encoded for each one.

\textit{Structure.} Section~\ref{sec:conceptual-model} presents \model, which is assessed in Section~\ref{sec:literature-assessment}. Section~\ref{sec:conclusion} concludes the paper.

\section{Conceptual Model: \model}
\label{sec:conceptual-model}

This section presents \model, the SE-oriented conceptual model of QC. We specify it using Unified Modeling Language (UML) class diagrams, with each diagram presenting one connected view of the concepts and relationships. Its goal is not to redefine established quantum notions, but to reorganize them into a structure that is easier to use for QSE.

\subsection{Scope and Organization of \model}

\model aims to help the QSE community understand QC systematically without requiring intensive study of quantum physics or mathematical formalism. First, concepts from the mathematical and physical foundations of QC are included only when they influence SE activities and their empirical settings. Second, QC is modeled in its SE context, so that quantum concepts are related to programs, computations, executions, deployment configurations, backends, and result evidence rather than treated as isolated mathematical or physical notions.

The rest of this section introduces \model incrementally, following the path from an application need to computation design, execution, and deployment. The application and software-system view (Section~\ref{sec:app-and-sys-view}) begins with familiar SE concerns and explains how an application need is formulated as a computational problem, constrained by requirements, addressed by a selected algorithm, and implemented in software. The computation-paradigm view (Section~\ref{subsec:quantum-computation-paradigm-view}) then explains that quantum computations can be realized in different ways, named computation paradigms, and introduces how a computation is described in each. Because current QSE studies primarily address gate-based computation, in which a computation is expressed as a sequence of discrete steps, the gate-based computation view (Section~\ref{subsec:gate-based-computation-view}) examines this form in greater detail. The execution and evolution view (Section~\ref{subsec:execution-evolution-view}) next distinguishes the intended computation from a concrete run, what happens during that run, and the results observed. The deployment view (Section~\ref{subsec:hardware-deployment-view}) explains how a computation is prepared for a simulator or physical quantum computer and how platform conditions affect its execution. Finally, the application examples (Section~\ref{subsec:application-examples}) connect these views in two worked cases. Note that like many SE artifacts, \model is incomplete by design and should evolve with QC technologies and QSE practices.

\subsection{Application and Software System View}
\label{sec:app-and-sys-view}

\begin{figure}[t]
\centering
\includegraphics[width=.8\columnwidth]{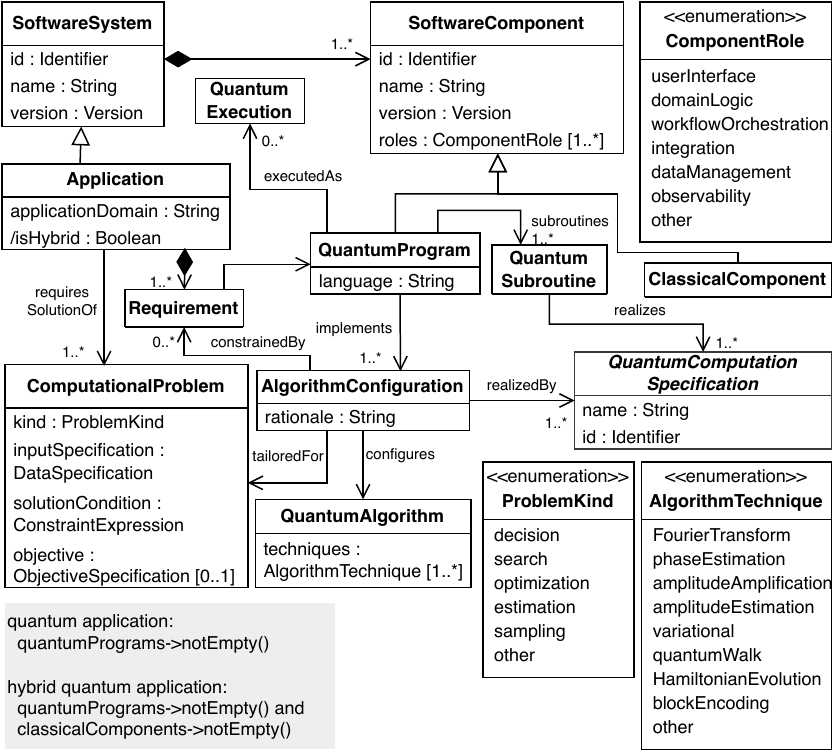}
\caption{Application and software-system view. The diagram connects requirements and computational problems to algorithm configurations, quantum programs and their subroutines, software components, and execution relations.}
\label{fig:application-stack-uml}
\end{figure}

Figure~\ref{fig:application-stack-uml} positions a quantum application as a software system that uses QC for solving application problems such as optimization and simulation~\cite{murillo2025qseroadmap}. In \model, an application is \emph{``quantum''} when its component structure includes at least one \texttt{QuantumProgram}, which subsequently contains one or more \texttt{QuantumSubroutine} instances, each representing a reusable portion of quantum computation that the program can invoke. Classical components may prepare inputs, coordinate workflows, invoke quantum computations, interpret results, or integrate the application with external systems.

An \texttt{Application} requires solutions to address one or more \texttt{ComputationalProblem} instances. Identifying the computational problem is critical in QSE because it is the point at which an application need becomes a computation-level contract. If this contract is vague, later choices about data encoding, circuit or Hamiltonian formulation, backend suitability, etc. become underspecified. In \model, a \texttt{ComputationalProblem} records the problem kind, expected input, validity condition for solutions, and, where relevant, an optimization or estimation objective.

For SE readers, \texttt{ProblemKind} is a property of a \texttt{ComputationalProblem} classifying its input-output contract; it is settled before the later algorithm, encoding, and paradigm choices, which it helps narrow without determining. Five literals are mutually exclusive because they are distinguished by output type: a Boolean (\texttt{decision}), any satisfying artifact (\texttt{search}), the artifact optimizing a stated objective (\texttt{optimization}), a value within a specified error (\texttt{estimation}), or a set of instances drawn from a target distribution (\texttt{sampling}). A sixth literal, \texttt{other}, is reserved for future extension.
A \texttt{decision} problem asks whether a given input satisfies a Boolean property, which produces a yes-or-no answer such as checking whether a feature configuration in a product line context is valid~\cite{eichhorn2023quantumcompu}.
A \texttt{search} problem returns any artifact satisfying a condition, such as finding a program satisfying a specification~\cite{hall2009aquantumalgo}.
An \texttt{optimization} problem returns a feasible artifact that optimizes a stated objective, such as a subset of a test suite~\cite{wang2024quantumappro}.
An \texttt{estimation} problem aims to approximate a numerical value within a specified error or confidence requirement. For instance, in hybrid QC workflows, repeated quantum executions produce measurement samples that are classically aggregated to estimate quantities such as Pauli-string expectation values~\cite{muqeet2024qops}.
A \texttt{sampling} problem aims to generate instances from a target distribution, rather than finding any valid instance as in \texttt{search} or the best instance as in \texttt{optimization}.
Note that a simulation or learning problem, common in QSE, is covered by one of these literals according to its deliverable: simulation, for example through Hamiltonian simulation \cite{peng2024simuq}, is an \texttt{estimation} problem whose target is a property of the modeled system; a learning problem, such as a hybrid quantum-machine-learning test oracle that learns expected robot behavior from historical data for regression testing~\cite{wang2025qmltestoracle}, is a \texttt{decision} or \texttt{estimation} problem according to whether its output is a label or a value.

The application also contains one or more \texttt{Requirement} instances. Requirements are critical in QSE because feasibility and quality must often be reasoned about early using quantum-specific constraints, such as available qubits, backend access, noise, execution budget, and acceptable failure probability \cite{yue2023towardsquant}. \model uses \texttt{Requirement} only as a bridge from application needs to \texttt{AlgorithmConfiguration}; it does not remodel requirements engineering concepts already established in SE.

A \texttt{QuantumAlgorithm} represents a reusable quantum solution approach that is independent of a particular application or execution environment. It can employ one or more algorithmic techniques or building blocks such as the quantum Fourier transform (QFT), phase estimation, amplitude amplification, amplitude estimation, variational optimization, quantum walks, or Hamiltonian evolution~\cite{nielsen2010quantum,montanaro2016quantum}.

An \texttt{AlgorithmConfiguration} represents the selection and application-specific tailoring of such an algorithm. It connects a general \texttt{QuantumAlgorithm} to the \texttt{ComputationalProblem} for which it is selected and to the \texttt{Requirement} instances that constrain the selection. Its rationale records why the algorithm and its particular configuration were considered appropriate. The configuration may capture decisions concerning algorithm parameters, encoding, oracle construction, error tolerance, classical--quantum partitioning, and resource limits. For example, the quantum approximate optimization algorithm (QAOA) is a reusable variational method; in IGDec-QAOA for test case optimization, the configuration includes the optimization objective encoding, \mbox{quadratic unconstrained binary optimization} (QUBO) formulation, decomposition strategy, circuit-depth, and backend or noise assumptions \cite{wang2024quantumappro}.
The distinction between \texttt{QuantumAlgorithm} and \texttt{AlgorithmConfiguration} is important for QSE: the former represents a reusable method with abstraction, whereas the latter helps instantiate that method into a concrete solution and documents how and why it is tailored to a specific engineering context.

An \texttt{AlgorithmConfiguration} is realized by one or more \texttt{QuantumComputation\allowbreak{}Specification} (\qcs) instances, which give the configured computation its paradigm-specific form rather than assuming a gate-based circuit. For instance, IGDec-QAOA realizes test case optimization through gate-based QAOA circuits \cite{wang2024quantumappro}, whereas BootQA realizes test case minimization by quantum annealing \cite{wang2024testcasemini}. These illustrate different configurations. A single configuration may still have several specifications when the selected algorithm, problem encoding, and constraints remain fixed. For example, the same QAOA configuration with a fixed QUBO encoding and circuit depth may be represented by multiple equivalent logical circuits.

Relying on the degree of code abstraction, a \texttt{QuantumProgram} can implement one or more \texttt{AlgorithmConfiguration} instances and construct or invoke one or more \qcs instances.
Reminding that the algorithm configuration captures the selected and tailored solution strategy, whereas the \qcs captures the concrete quantum computation that realizes that strategy, a quantum program provides the executable software that connects the two concepts to the remainder of the application. More specifically, the quantum program may prepare input data, bind parameters, construct circuits or energy models, submit computations to a backend, coordinate classical feedback, and process the resulting classical outputs. A \texttt{QuantumProgram} may be executed zero or more times: it may remain defined without being invoked, or it may have multiple runtime executions, each represented by a \texttt{QuantumExecution}. This separates the reusable software definition from its concrete runtime occurrences and allows programs to be modeled before execution. The runtime details of execution and results are discussed in the execution and evolution view (Section~\ref{subsec:execution-evolution-view}).

Overall, this view establishes the abstraction path followed by \model: an application need is expressed as a computational problem and constrained by requirements; an algorithm is then selected and configured; the configuration is realized as a paradigm-specific computation specification; and executable programs connect that specification to the surrounding software system. This keeps quantum application engineering grounded in familiar SE abstractions while making explicit where QC-specific decisions enter. The following views refine the computation, operation, execution, evolution, hardware, and deployment concepts introduced here.

\subsection{Quantum Computation Paradigm View}
\label{subsec:quantum-computation-paradigm-view}

\begin{figure}[t]
\centering
\includegraphics[width=.8\columnwidth]{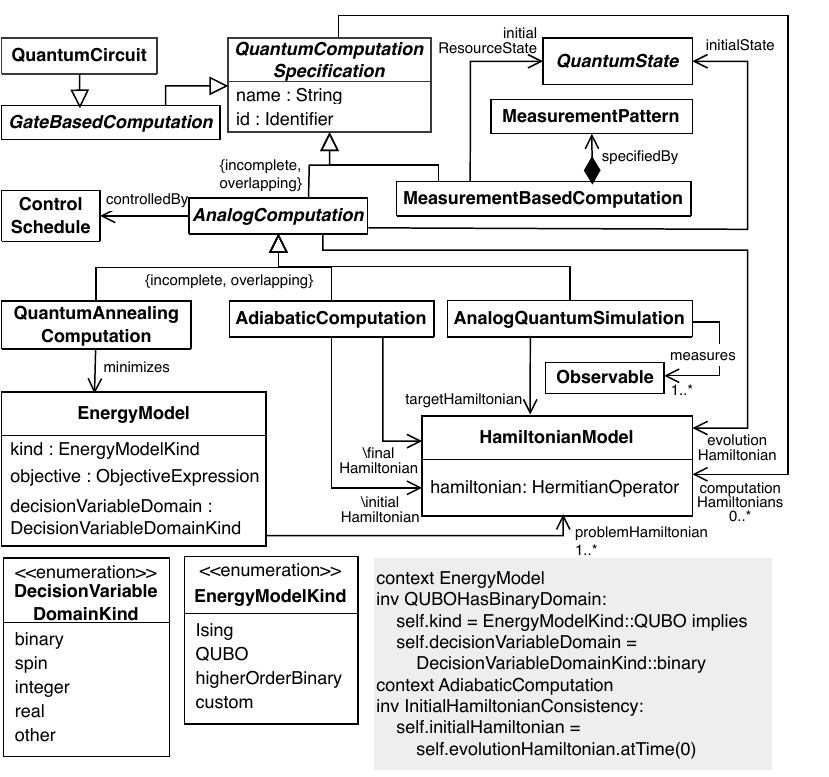}
\caption{Quantum-computation paradigm view. The diagram represents computation specifications as circuits, Hamiltonian models with control schedules, energy models, or measurement patterns, with illustrative well-formedness constraints.}
\label{fig:alternative-paradigms-uml}
\end{figure}

As shown in Figure~\ref{fig:alternative-paradigms-uml}, a \qcs specifies the paradigm-specific structure of the quantum computation used to realize the algorithm selected and tailored by an \texttt{AlgorithmConfiguration}. Different quantum computation paradigms specify computation through different metaclasses, e.g., \texttt{GateBasedComputation} is represented by a \texttt{QuantumCircuit}~\cite{nielsen2010quantum}; \texttt{AnalogComputation} by a \texttt{HamiltonianModel} and \texttt{ControlSchedule}~\cite{lloyd1996universal}; \texttt{QuantumAnnealingComputation} by an \texttt{EnergyModel}~\cite{kadowaki1998quantum}; and \texttt{MeasurementBasedComputation} by a \texttt{MeasurementPattern} for measurement-based quantum computation (MBQC)~\cite{raussendorf2001oneway}.
\model treats the paradigm hierarchy as incomplete and overlapping: new paradigms can be added. Well-known platforms and software ecosystems illustrate the distinction: IBM Quantum provides access to circuit-based QC services~\cite{ibm2026qiskit}; D-Wave systems expose annealing over Ising or QUBO models~\cite{dwave2026annealing}, QuEra's Aquila supports analog Hamiltonian simulation~\cite{aws2026quera}, and Xanadu describes a photonic MBQC architecture~\cite{xanadu2026photonics}.

A \texttt{GateBasedComputation} specifies a computation via discrete quantum operations organized as a \texttt{QuantumCircuit} \cite{nielsen2010quantum}. Because circuit-based abstractions appear frequently in the QSE literature, their qubits, states, operations, and circuit structure are discussed separately in Section~\ref{subsec:gate-based-computation-view}.

An \texttt{AnalogComputation} specifies a computation through the continuous-time evolution of a quantum system under a possibly time-dependent Hamiltonian, a mathematical operator that describes the system's energy structure and governs how its quantum state evolves over time. A \texttt{HamiltonianModel} is a software-level representation of a mathematical Hamiltonian. It records the Hamiltonian's terms, coefficients, parameters, and, where applicable, time dependence, so that the Hamiltonian can be constructed, transformed, validated, simulated, and adapted for deployment. A \texttt{ControlSchedule} specifies how externally controllable parameters of the Hamiltonian vary over time. Together, the Hamiltonian model and control schedule determine the intended evolution, and an initial \texttt{QuantumState} provides its starting condition.

A \texttt{QuantumAnnealingComputation} is an optimization subtype of \texttt{AnalogComputation}. It applies when an application-level \texttt{ComputationalProblem} has \texttt{ProblemKind::optimization} (Figure~\ref{fig:application-stack-uml}) and an \texttt{AlgorithmConfiguration} selects annealing as the strategy. In \model, the \texttt{EnergyModel} is the optimization artifact: it states the objective over \texttt{DecisionVariable} instances as an Ising, QUBO, higher-order binary, etc. QUBO and higher-order binary models use variables in $\{0,1\}$, whereas Ising models use spins in $\{-1,+1\}$. Application-level variables with other variables such as integer, real-valued, or categorical variables, must first be mapped to binary variables or Ising spins, with any necessary constraints incorporated into the model. For execution, the energy model is represented as a problem Hamiltonian whose low-energy states correspond to better solutions and whose ground states represent optima. The annealing process is still analog: the full time-dependent evolution Hamiltonian includes this problem Hamiltonian and is driven by a \texttt{ControlSchedule} toward low-energy states \cite{kadowaki1998quantum}. In QSE, this modeling approach is used when test case minimization is encoded as a QUBO for quantum annealers \cite{wang2024testcasemini}, when test optimization is formulated as Ising models for Ising-based solvers and benchmarks \cite{yang2026isingtest}, and when dynamic analysis of D-Wave programs recovers Hamiltonian expressions representing the underlying optimization problem definitions \cite{perezcastillo2023dynamic}.

An \texttt{AdiabaticComputation} is an \texttt{AnalogComputation} whose Hamiltonian changes gradually from an initial to a final Hamiltonian according to a \texttt{ControlSchedule} \cite{farhi2001quantum}. The initial Hamiltonian is chosen so that its relevant low-energy state can be prepared, while the final Hamiltonian encodes the intended computational result. Under the adiabatic condition, the system remains close to the corresponding instantaneous eigenstate during the evolution. A \texttt{QuantumAnnealingComputation} may use this adiabatic principle when its annealing schedule slowly moves from a driver Hamiltonian to the problem Hamiltonian.

An \texttt{AnalogQuantumSimulation} uses an engineered quantum system to reproduce or approximate another, target system whose behavior is impractical to obtain through direct experiments or classical simulation at the required scale, such as a molecule or magnetic material~\cite{lloyd1996universal}. Its \texttt{targetHamiltonian} describes the closed-system model to be simulated; the computation prepares an initial state, configures an implemented Hamiltonian, evolves the system, and measures \texttt{Observable} instances, i.e., properties of the simulated system that the application wants to estimate. Examples include the system's energy, the number of particles or excitations, or correlations between parts of the system. The implemented Hamiltonian may equal the target Hamiltonian in a direct simulation or approximate it through an experimentally realizable mapping. This distinction between target and implemented Hamiltonians is reflected in software support for analog quantum simulation. For example, SimuQ lets users specify a target Hamiltonian, such as an Ising spin model, and compile it to realizations on devices such as neutral-atom arrays, superconducting transmon systems, and trapped-ion systems \cite{peng2024simuq}. Classical simulators provide related but distinct support: QuTiP's pulse-level engine numerically integrates Hamiltonians, control pulses, and noise models on a conventional computer to emulate hardware-aware circuit behavior \cite{li2022pulselevelnoisy}.

In MBQC, a computation prepares an entangled resource state and then proceeds through a sequence of measurements. This should not be confused with measurement outcomes or records used as observations in SE tasks such as testing and debugging \cite{garcia2023qst,wang2021qdiffdiffere}. In \model, a \texttt{MeasurementBasedComputation} captures this paradigm through a \texttt{MeasurementPattern} \cite{raussendorf2001oneway}. MBQC has been demonstrated with photonic cluster states and trapped ions \cite{walther2005experimental,lanyon2013measurement}. The pattern is the main engineering artifact: it specifies which subsystems are measured, which bases or angles are used, how later measurements depend on earlier outcomes, and which classical feed-forward corrections are applied. Measurement is therefore part of the computation, not only its final readout, which changes what must be specified, tested, and verified.

These paradigms also lead to different engineering artifacts. Gate-based specifications are refined through circuits, native-operation sets, and qubit mappings; annealing specifications through energy models and embeddings; analog specifications through Hamiltonians and control schedules; and MBQC specifications through resource states and adaptive measurement patterns. \model therefore treats \qcs as an extensible abstraction so that QSE methods can refer to paradigm-specific artifacts without assuming that every computation is ultimately circuit-based.

\subsection{Gate-Based Computation View}
\label{subsec:gate-based-computation-view}

Gate-based computation is currently the most extensively represented form of QC in QSE, in modeling and code-generation tools \cite{gemeinhardt2021towardsmodel,guo2025m2qcodeamode} and especially in quantum software and program testing \cite{garcia2023qst,fortunato2024vv,mendiluze2021muskitamutat,paltenghi2023morphqmetamo,wang2021qdiffdiffere,paltenghi2024analyzingqua,oldfield2025qst}. This focus also reflects the broader QC software stack that QSE methods often target: OpenQASM provides a circuit-oriented assembly language and intermediate representation \cite{cross2022openqasm3}, while Qiskit/IBM Quantum supports circuit construction, transpilation, and backend execution \cite{ibm2026qiskit}. \model therefore describes gate-based computation in greater detail than the other computation paradigms, reflecting the present focus of QSE research and practice.

\subsubsection{Circuit and Operation View}
\label{subsubsec:circuit-operation-view}

\begin{figure}[t]
\centering
\includegraphics[width=.8\columnwidth]{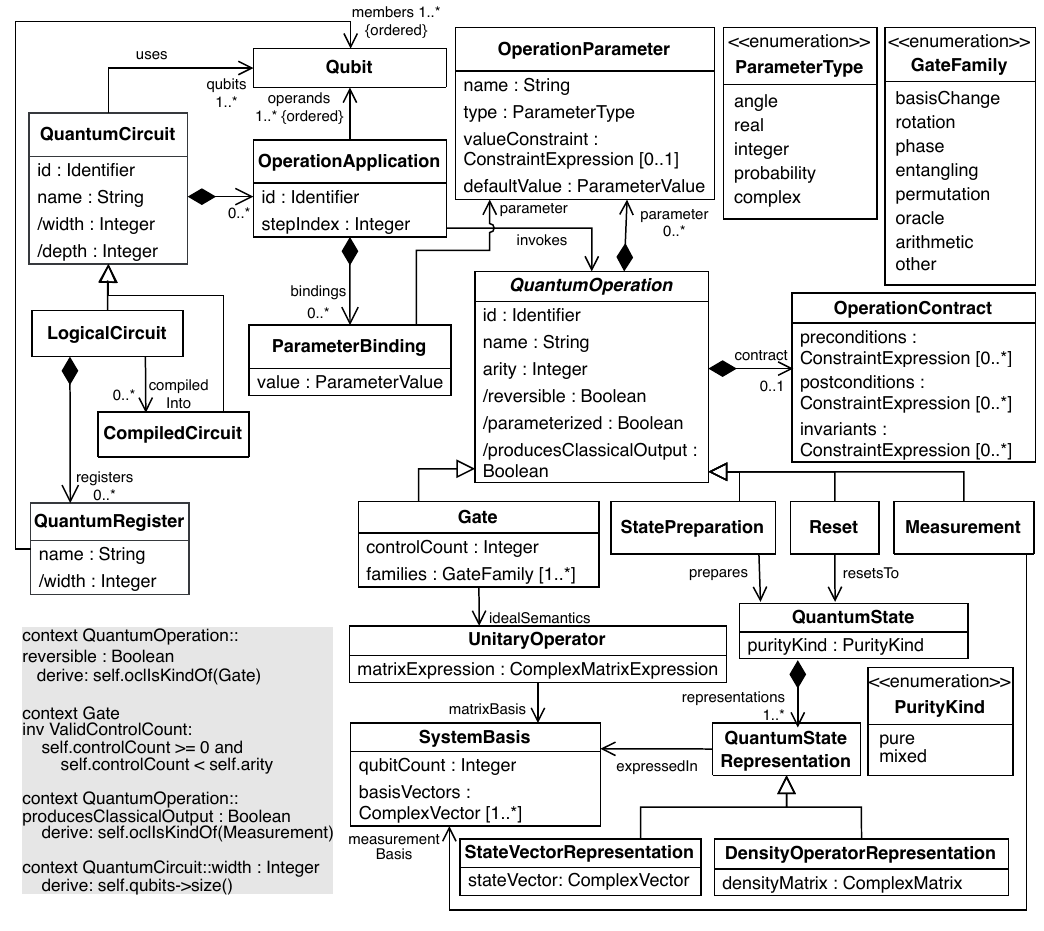}
\caption{Gate-based operation view. The diagram separates reusable operation definitions from circuit applications and relates operations to operands, parameters, contracts, states, representations, and bases.}
\label{fig:operation-model-uml}
\end{figure}

As shown in Figure~\ref{fig:operation-model-uml}, a \texttt{QuantumCircuit} represents gate-based computation via \texttt{QuantumRegister} instances that group its \texttt{Qubit} instances. The circuit's width is derived from the total number of qubits across its registers, while its depth is derived from the \texttt{stepIndex} values assigned to operation applications. For SE, width and depth affect resource estimation, backend feasibility, exposure to noise, testing, and optimization. Operations sharing a \texttt{stepIndex} are only candidates for parallel execution: they may still be serialized if they use the same qubit or if backend constraints prevent parallel execution.
\model distinguishes \texttt{LogicalCircuit} and \texttt{CompiledCircuit}. A logical circuit captures the backend-independent computation, whereas a compiled circuit records the backend-adapted realization produced by compilation. Physical placement is represented by \texttt{QubitMapping} instances in the gate-based deployment artifact discussed in the deployment view (Section~\ref{subsec:hardware-deployment-view}). This distinction matters in SE: it separates circuit-level operands from deployment-specific placement, letting tests, analyses, and defects target either the intended circuit or its deployed realization.

A \texttt{QuantumCircuit} contains zero or more \texttt{OperationApplication} instances. Each operation application is one occurrence of a \texttt{QuantumOperation} in the circuit: it records the selected operation, its ordered qubit operands, its \texttt{stepIndex}, and any \texttt{ParameterBinding} instances used in that occurrence.
The arity of a \texttt{QuantumOperation} specifies how many operands an application must provide. Each \texttt{OperationApplication} must supply exactly that number of distinct qubits, and each of those qubits must belong to the containing circuit. The order is semantically relevant: for a controlled-NOT (CNOT) gate, the first operand may denote the control qubit and the second the target qubit, so exchanging them changes the operation. These conditions ensure that each operation application conforms to the invoked operation.

A \texttt{QuantumOperation} may own \texttt{OperationParameter} instances, each with a name, a type, an optional \texttt{valueConstraint}, and an optional default. Supported types include the real, integer, and complex numeric domains, as well as semantically constrained reals such as rotation angles (periodic modulo $2\pi$) and probabilities (bounded in $[0,1]$).
A \texttt{ParameterBinding} assigns a concrete value to an \texttt{OperationParameter} at a particular \texttt{OperationApplication}. The distinction is needed because the parameter belongs to the reusable operation definition, while the value belongs to one configured occurrence in a circuit. For example, the same $R_z(\theta)$ operation can appear several times with different $\theta$ values. Each binding must therefore refer to a parameter owned by the invoked operation, and an operation application should not bind the same parameter more than once.
A \texttt{QuantumOperation} may also own an optional \texttt{OperationContract} that records preconditions, postconditions, and invariants, providing an operation-level place to reference assumptions and expected properties used in specification-based testing and test oracles \cite{oldfield2025qst,li2026dynamicoracle} and runtime assertions \cite{rovara2025runtimeassertions}.

\model distinguishes four kinds of state-affecting quantum operation: \texttt{Gate}, \texttt{StatePreparation}, \texttt{Reset}, and \texttt{Measurement}. They cover the gate-based operation kinds needed here: unitary change, state initialization, state reinitialization, and observation with possible state update.
A \texttt{Gate} represents a reversible operation whose semantics is given by one \texttt{UnitaryOperator}; for arity $n$, the unitary matrix has dimension $2^n \times 2^n$, so \texttt{reversible} holds. A gate has one or more \texttt{GateFamily} values, such as basis-change, rotation, phase, entangling, permutation, oracle, or arithmetic. The families are not exclusive: the Hadamard gate can be treated as basis-changing, $R_z(\theta)$ as a rotation and phase gate, and CNOT as an entangling controlled gate. The \texttt{controlCount} records the number of control operands and must be smaller than the gate arity.
A \texttt{StatePreparation} establishes a designated \texttt{QuantumState}: a single-qubit state such as $|0\rangle$ or $|+\rangle=(|0\rangle+|1\rangle)/\sqrt{2}$, or a possibly multi-qubit, application-specific input state. A \texttt{Reset} reinitializes an operand to a target \texttt{QuantumState}, commonly $|0\rangle$, so it can be reused or measured again. \model separates them from gates because they establish state rather than apply a unitary to an operand, and separates reset from state preparation because QSE analyses distinguish input construction from execution-time reset behavior \cite{garcia2023qst,ma2025qmon}.
A \texttt{Measurement} is specified relative to one \texttt{SystemBasis} and is the only operation kind for which \texttt{producesClassicalOutput} holds. It is modeled separately from gates because it is non-unitary, and because the chosen basis determines how quantum state information becomes classical observations. In QSE, measurements link quantum executions to classical evidence.

A \texttt{QuantumState} records whether the state is pure or mixed through \texttt{PurityKind} and may be associated with one or more \texttt{QuantumStateRepresentation} instances. The purity classification separates what kind of state is described from the mathematical form used to express it. For QSE, pure and mixed states need different forms: pure states are expressed as state vectors and suffice for ideal simulation and circuit-level equivalence checking, whereas mixed states are expressed as density operators and are needed for noise, measurement uncertainty, or partial information. Because the same vector or matrix denotes different states under different bases, a representation means nothing without the \texttt{SystemBasis} it was written in.
A \texttt{SystemBasis} specifies the number of qubits and the ordered basis vectors used to express states, operators, and measurement outcomes. For $n$ qubits, a complete basis contains $2^n$ mutually orthonormal vectors, each of dimension $2^n$. Testing work depends on the measurement basis, for example when projective tests use computational-basis measurements \cite{oldfield2025qst} or when test oracles assume a fixed basis for interpreting outcomes \cite{li2026dynamicoracle}.
A \texttt{StateVectorRepresentation} expresses a pure state of an $n$-qubit system as a normalized complex vector with $2^n$ components. A \texttt{DensityOperatorRepresentation} expresses a state as a $2^n \times 2^n$ complex matrix and can represent both pure and mixed states. The OCL constraints in \model are a subset that targets conceptual-model consistency, such as circuit width, measurement output, and control-count validity, leaving standard mathematical constraints such as vector normalization and operator unitarity implicit.

Overall, the gate-based view connects gate-based computation to familiar SE concerns such as reuse, parametrization, contracts, and transformation, representing each through model constructs such as operation definitions, invocation occurrences, and parameter bindings. These distinctions support traceability from logical circuit specification through compilation, execution, and testing.

\subsubsection{Qubit Characterization View}
\label{subsubsec:qubit-characterization-view}

\begin{figure}[t]
\centering
\includegraphics[width=.8\columnwidth]{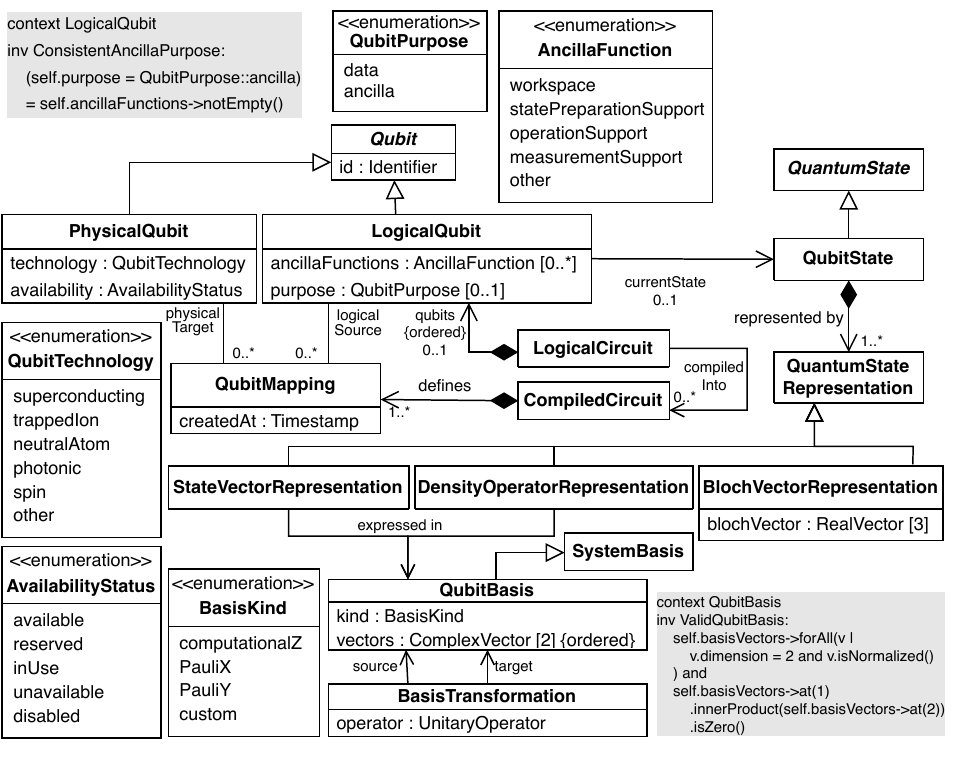}
\caption{Qubit-characterization view. The diagram distinguishes logical and physical qubits and links qubit roles, states, representations, bases, and deployment-relevant physical attributes.}
\label{fig:qubit-uml}
\end{figure}
Figure~\ref{fig:qubit-uml} separates qubits used in the logical circuit from qubits provided by hardware. A \texttt{LogicalCircuit} consists of \texttt{LogicalQubit} instances that serve as circuit operands, while a \texttt{PhysicalQubit} is a qubit realized in hardware via a \texttt{QubitTechnology}, e.g., superconducting, trapped-ion, neutral-atom, photonic, or spin. This distinction is central for QSE because a program and its tests are usually written over logical qubits, whereas execution depends on the physical qubits selected during compilation and deployment.

\model classifies logical qubits by \texttt{QubitPurpose}. A data qubit carries application or intermediate data, while an ancilla qubit supports auxiliary computation, state preparation, or measurement. The OCL constraint \texttt{ConsistentAncillaPurpose} requires an ancilla to have at least one \texttt{AncillaFunction} and prevents data qubits from having ancilla functions. Role-dependent expectations matter in testing: for example, a temporary-computation ancilla must return to a known state before reuse, while a measurement-support ancilla is measured to obtain test evidence through projective measurements \cite{oldfield2025qst} or runtime assertions \cite{rovara2025runtimeassertions}.
A \texttt{PhysicalQubit} has a \texttt{QubitTechnology} and an \texttt{availability} value capturing deployment-relevant facts: a physical qubit may exist on a processor but be unavailable or unsuitable for a particular execution. Together with calibration and topology information (Section~\ref{subsec:hardware-deployment-view}), these properties determine whether a compiled circuit can execute on a selected backend. Additional characteristics such as coherence, readout fidelity, and operation quality are not attributes of \texttt{PhysicalQubit}; they are recorded through the calibration context (Section~\ref{subsec:hardware-deployment-view}), because \model focuses on SE-level deployment reasoning rather than detailed hardware characterization.

\subsubsection{Qubit State and Basis View}
\label{subsubsec:qubit-state-view}
Figure~\ref{fig:qubit-uml} shows the local state of an individual qubit. A \texttt{QubitState} represents the local state of an individual \texttt{LogicalQubit}, while \texttt{QuantumSystemState} (Section~\ref{subsec:execution-evolution-view}) represents the joint state of multiple qubits. This separation supports QSE tasks that require qubit-level expectations for initialization, ancilla reuse, measurement support, and local assertions: a single-qubit assertion refers to a \texttt{QubitState}, whereas a simulator snapshot, an entangled circuit state, or a test oracle checking joint-state properties refers to a \texttt{QuantumSystemState}.

\texttt{QubitState} inherits the pure/mixed distinction introduced for \texttt{QuantumState}. This distinction matters for local states because a qubit inside an entangled multi-qubit system may have a mixed local state even when the joint \texttt{QuantumSystemState} is pure, and because noise induces mixed states that testing must represent \cite{muqeet2024mitigatingno}.
A \texttt{QubitState} can use three representation forms. A \texttt{StateVectorRepresentation} records a two-component vector and is valid when the local qubit state is pure. A \texttt{DensityOperatorRepresentation} records a $2 \times 2$ density matrix and can express both pure and mixed local states, including reduced states arising from entanglement or noise. A \texttt{BlochVectorRepresentation} records a three-dimensional real vector for compact single-qubit visualization or assertion-style reasoning. These representations support local qubit-state expectations such as initialization conditions and runtime state assertions \cite{oldfield2025qst,rovara2025runtimeassertions}.

\texttt{QubitBasis} specializes \texttt{SystemBasis} (Section~\ref{subsubsec:circuit-operation-view}), applying it to an individual \texttt{QubitState} rather than an entire quantum system. Basis choices affect measurement interpretation, test oracles, and debugging \cite{oldfield2025qst,li2026dynamicoracle}. For example, Pauli-string testing requires measuring each qubit in the Pauli eigenbasis corresponding to its operator ($X$, $Y$, or $Z$)~\cite{muqeet2024qops}, and projection-based assertions may require basis transformations to accommodate measurement-restricted hardware~\cite{li2020proq}. A \texttt{BasisTransformation} represents such a change of coordinates between qubit bases, realized through its \texttt{operator} attribute, a unitary operator.

\subsection{Execution and Evolution View}
\label{subsec:execution-evolution-view}

\begin{figure}[t]
\centering
\includegraphics[width=.8\columnwidth]{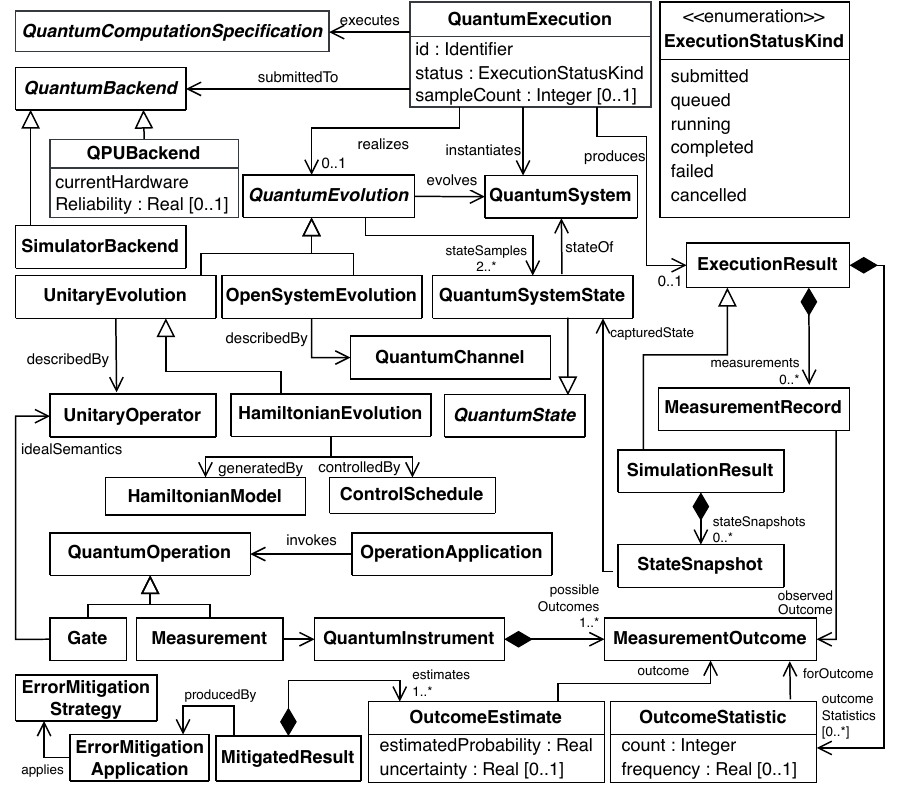}
\caption{Execution and evolution view. The diagram relates computation specifications to executions, sample counts, backends, logical quantum systems, evolution models, measurements, simulation outputs, and execution results, including derived estimates and mitigated results.}
\label{fig:execution-evolution-uml}
\end{figure}

Figure~\ref{fig:execution-evolution-uml} refines the execution relation of Figure~\ref{fig:application-stack-uml} to cover runtime evidence. A \qcs is executed by a \texttt{QuantumExecution}, which is submitted to one \texttt{QuantumBackend}, carries an \texttt{ExecutionStatusKind}, instantiates a \texttt{QuantumSystem}, and may realize at most one \texttt{QuantumEvolution} and produce at most one \texttt{ExecutionResult}.
Here, \texttt{QuantumSystem} denotes the logical, state-bearing system considered during that run, not the backend. In this view, each element plays a distinct role: the specification states the intended computation, the backend identifies where it is run, the quantum system defines the logical state being evolved, and the execution result records the raw evidence from that run. Derived estimates and error-mitigated results are represented separately.

\subsubsection{Quantum Execution}
\label{subsubsec:quantum-execution}
A \texttt{QuantumExecution} represents one concrete attempt to run a \qcs. The class is needed because a specification can be executed repeatedly under different execution conditions, such as parameter bindings, sample count, backend choice, noise assumptions, solver settings, or run repetitions. The \texttt{sampleCount} attribute records how many repeated observations or runs are requested for that execution. Gate-based systems commonly call these repetitions shots, while quantum annealers call them reads. For example, the same gate-based circuit, annealing problem, or analog simulation specification may be evaluated in different execution environments or with different \texttt{sampleCount} values. These runs are not interchangeable evidence: QSE studies must record the execution context to interpret and reproduce results \cite{mauerer2022123reproduci}. Quantum-optimization studies for SE similarly need to distinguish the formulated optimization problem from the solver or execution target used to evaluate it and report settings such as repetitions, shots, noise handling, baselines, and metrics \cite{zhang2026quantumoptimization,zhang2025empiricalstudies}.

Each execution is submitted to one \texttt{QuantumBackend} and has an \texttt{ExecutionStatusKind}. \model distinguishes \texttt{QPUBackend} from \texttt{SimulatorBackend}, a distinction reflected in QSE testing studies that compare behavior across quantum software stacks and backend configurations \cite{wang2021qdiffdiffere,paltenghi2023morphqmetamo,muqeet2024mitigatingno}. The status values in Figure~\ref{fig:execution-evolution-uml} record whether the job is submitted, queued, running, completed, failed, or cancelled. Together, backend and status provide the minimum context for monitoring, retry policies, failure diagnosis, and reproducibility records.

An execution may produce zero or one \texttt{ExecutionResult}, which may contain raw \texttt{MeasurementRecord} instances, each linked to the \texttt{MeasurementOutcome} it observes. Outcomes are possible results; records are observations. \model distinguishes three quantities. A \texttt{Measurement} is associated with a \texttt{QuantumInstrument} that specifies outcomes, probabilities, and state updates \cite{nielsen2010quantum}; these probabilities give the \emph{expected} distribution over \texttt{MeasurementOutcome}s. The \emph{observed} \texttt{OutcomeStatistic} summarizes the records using a count or frequency. The \emph{inferred} \texttt{OutcomeEstimate} records a derived estimate, such as an estimated probability and its uncertainty. An \texttt{ErrorMitigationApplication} applies an \texttt{ErrorMitigationStrategy} to raw execution evidence and produces a \texttt{MitigatedResult} alongside it, augmenting rather than replacing the raw observations. These distinctions support statistical testing, where counts, frequencies, and estimates are interpreted stochastically rather than as deterministic returns \cite{fortunato2024vv,sato2025buglocating}.
Simulator-specific evidence is represented as a distinct subtype. A \texttt{SimulationResult} specializes \texttt{ExecutionResult} and may contain \texttt{StateSnapshot} instances. Each snapshot captures a \texttt{QuantumSystemState} at a selected point, such as a saved statevector for a circuit simulation or a sampled state representation for an analog simulation. Such snapshots are simulator-side artifacts, not ordinary measurement records: they should not be treated as directly observed intermediate states, since accessing an unknown quantum state generally requires additional measurement procedures \cite{nielsen2010quantum}.

\subsubsection{Quantum Evolution}
\label{subsubsec:quantum-evolution}
A \texttt{QuantumEvolution} describes how the \texttt{QuantumSystemState} of the logical \texttt{QuantumSystem} changes during an execution. An execution may realize zero or one explicitly represented \texttt{QuantumEvolution}; the zero case covers executions whose evolution is unavailable, such as failed jobs or runs that expose only final observations. \texttt{QuantumSystemState} is a joint state, not a list of independent local states. This matters for SE because tests, assertions, debugging evidence, and simulator snapshots may need to reason about joint-state properties rather than checking each local subsystem independently, especially when entanglement is involved \cite{nielsen2010quantum,fortunato2024vv,rovara2025runtimeassertions}. The \texttt{stateSamples} association records selected states of an evolution, such as initial, intermediate, and final states. These samples are a finite engineering view for specification, simulation, testing, or visualization. They do not imply that continuous evolution is inherently discrete.

\model distinguishes ideal closed-system evolution from open-system evolution. \texttt{UnitaryEvolution} is described by a \texttt{UnitaryOperator} and captures reversible behavior, such as the ideal effect of a gate sequence before measurement or noise \cite{nielsen2010quantum}. \texttt{HamiltonianEvolution} is a unitary subtype whose dynamics is generated by a \texttt{HamiltonianModel}; a \texttt{ControlSchedule} specifies time-varying controls for that model when needed (see Section~\ref{subsec:quantum-computation-paradigm-view}) \cite{lloyd1996universal,kadowaki1998quantum,farhi2001quantum}. \texttt{OpenSystemEvolution} is described by a \texttt{QuantumChannel} and covers evolution affected by measurement, reset, decoherence, amplitude damping, depolarizing noise, or other interaction with an environment \cite{nielsen2010quantum}.

The distinction is practical for QSE. Unitary reasoning supports equivalence checking, reversible-transformation reasoning, and statevector simulation. Open-system reasoning is needed whenever non-ideal effects influence the observed result, including timing effects. A transformation that is sound under ideal unitary semantics may not preserve behavior under open-system or noisy execution semantics. Explicit evolution kinds therefore help engineers choose suitable simulators, test oracles, equivalence relations, and debugging explanations \cite{garcia2023qst,fortunato2024vv,wang2021qdiffdiffere}.

Overall, this view separates the intended computation, the execution attempt, the modeled evolution of the logical system, and the evidence returned. This gives QSE methods traceability from design intent to runtime evidence: what was specified, where and whether it ran, how its evolution was modeled, and what observations were obtained.

\subsection{Deployment View}
\label{subsec:hardware-deployment-view}
Deployment concepts of \model can be divided into paradigm-neutral elements including deployment configurations and artifacts, backends, calibration profiles, and individual execution occurrences, and paradigm-specific ones. For gate-based systems these include compiled circuits, native operations, device topologies, and qubit mappings; for analog and annealing systems, device Hamiltonians, physical subsystems, interaction topologies, and variable mappings. These concepts affect feasibility, portability, reproducibility, and cost \cite{yue2026qready}.

\subsubsection{Gate-Based Deployment}
\label{subsubsec:gate-based-deployment}

\begin{figure}[t]
\centering
\includegraphics[width=.8\columnwidth]{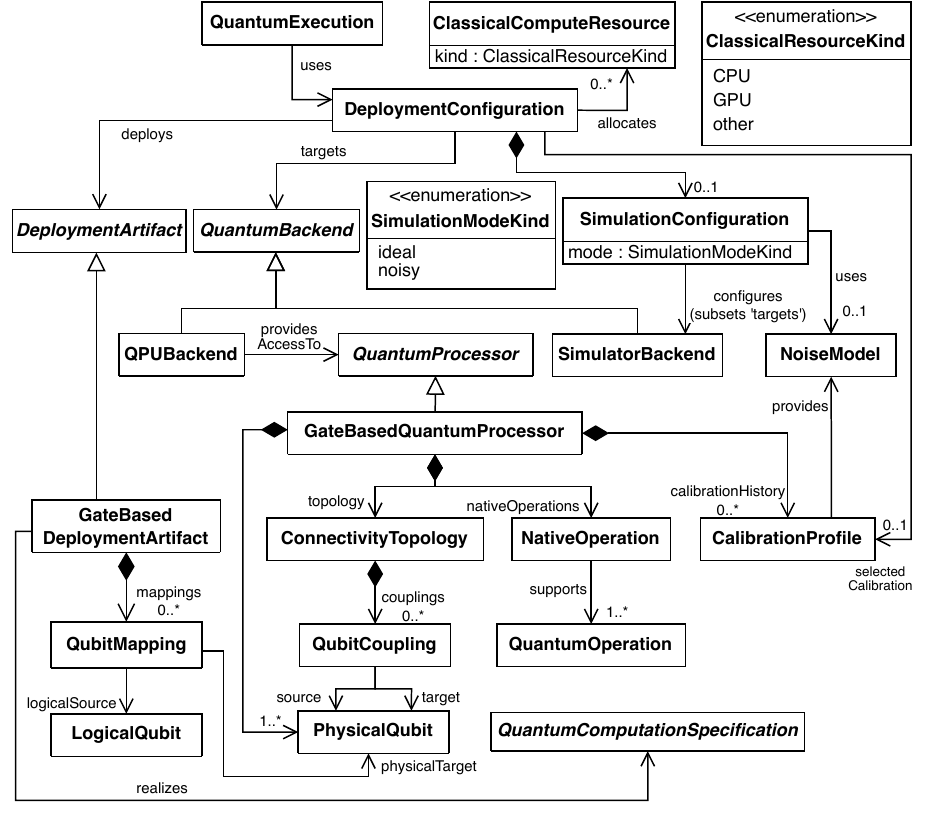}
\caption{Gate-based deployment view. The diagram relates executions to deployment configurations, artifacts, backends, noise models, processor capabilities, calibration profiles, and qubit mappings.}
\label{fig:deployment-gate-based-uml}
\end{figure}

As shown in Figure~\ref{fig:deployment-gate-based-uml}, each \texttt{QuantumExecution} uses a \texttt{DeploymentConfiguration}, which identifies a \texttt{DeploymentArtifact}, a target \texttt{QuantumBackend}, and execution context such as compiler or transpiler settings, software development kit (SDK) version, selected calibration profile, and, for simulator runs, a \texttt{SimulationConfiguration} and optional \texttt{NoiseModel}. This separation makes repeated executions, alternative deployment choices, and their shared computation specification traceable \cite{mauerer2022123reproduci}.

\model distinguishes \texttt{QPUBackend} and \texttt{SimulatorBackend} because QSE studies show that backend choice, stack behavior, noise assumptions, and hardware constraints can affect testing evidence, feasibility assessment, and reproducibility \cite{wang2021qdiffdiffere,paltenghi2023morphqmetamo,muqeet2024mitigatingno,yue2026qready}. For QPU execution, the target backend provides access to a \texttt{GateBasedQuantumProcessor}. \model records only deployment-relevant information rather than low-level hardware behavior: \texttt{ConnectivityTopology} constrains which qubits can interact directly; \texttt{NativeOperation} identifies operations supported by the processor; and \texttt{CalibrationProfile} records time-varying information such as error rates, gate duration, readout quality, and coherence.
Together, this information provides provenance for backend-dependent decisions.

A \texttt{GateBasedDeploymentArtifact} is the backend-specific realization of the \qcs. It may contain a compiled circuit, resolved parameters, decompositions, scheduling information, and one or more \texttt{QubitMapping} instances. Because a processor supports only particular qubits and native operations, compilation may introduce routing operations such as SWAPs (state exchanges between qubits) or decompose unsupported logical operations. Each \texttt{QubitMapping} connects a \texttt{LogicalQubit} to a \texttt{PhysicalQubit} of the selected processor, making physical placement explicit. This matters because tests and analyses are usually expressed over logical qubits, whereas QPU behavior is conditioned by the selected processor, topology, calibration state, and mapped physical qubits \cite{paltenghi2023morphqmetamo,muqeet2024mitigatingno}.

For simulator deployment, a \texttt{SimulationConfiguration} specifies whether execution uses ideal or noisy simulation and may configure the targeted \texttt{SimulatorBackend}. Ideal simulation supports reference results for testing and validation. Noisy simulation uses a \texttt{NoiseModel} to approximate selected deviations from ideal quantum evolution, such as operation errors, decoherence, state-preparation errors, or measurement errors \cite{nielsen2010quantum}. A calibration profile may inform a \texttt{NoiseModel}, but \model keeps \texttt{CalibrationProfile} and \texttt{NoiseModel} separate because QSE often uses noise models as execution assumptions for testing, simulation, feasibility assessment, or result interpretation, not as complete device descriptions \cite{wang2021qdiffdiffere,muqeet2024mitigatingno,mauerer2022123reproduci,yue2026qready}. A noise model can therefore be reused, simplified, or constructed independently of a particular calibration profile.

Overall, the gate-based deployment view links execution to its deployable artifact, target backend, physical mapping, calibration context, and simulation assumptions. QSE studies using gate-based algorithms, such as QAOA-based test case optimization, rely on backend-adapted circuits, parameter settings, and execution assumptions to interpret reported results \cite{wang2024quantumappro}. Testing, platform, noise-mitigation, and reproducibility studies further show that backend identity, transpilation, noise assumptions, and execution settings affect runtime evidence and diagnosis \cite{wang2021qdiffdiffere,paltenghi2023morphqmetamo,muqeet2024mitigatingno,mauerer2022123reproduci}.

\subsubsection{Analog and Quantum-Annealing Deployment}
\label{subsubsec:analog-annealing-execution}
\begin{figure}[t]
\centering
\includegraphics[width=0.7\columnwidth]{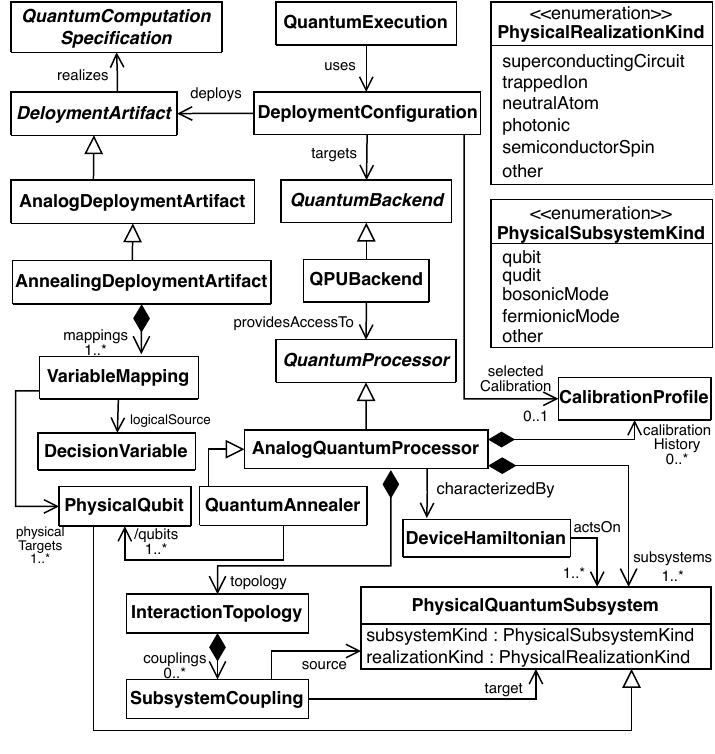}
\caption{Analog and quantum-annealing deployment view. The diagram extends deployment with analog artifacts, device Hamiltonians, physical subsystems, interaction topologies, calibration profiles, annealers, and variable mappings.}
\label{fig:analog-qc-deployment-uml}
\end{figure}

Figure~\ref{fig:analog-qc-deployment-uml} reuses the deployment concepts of the gate-based view and adds concepts specific to continuous evolution and device-level analog modeling.
As in gate-based deployment, an \texttt{AnalogDeploymentArtifact} is the deployable analog realization, such as hardware-compatible Hamiltonian parameters and a control schedule. The following paragraphs first define the backend context to interpret this realization---the selected device Hamiltonian, physical subsystems, interaction topology, and calibration profile---and then describe the annealing-specific evolution terms and how they are mapped to physical resources.

\textbf{Device characterization.}
An \texttt{AnalogQuantumProcessor} is mainly characterized by a \texttt{DeviceHamiltonian}: a mathematical model of the fixed and controllable physical dynamics available for execution. This is the analog counterpart to the gate-based characterization described in Section~\ref{subsubsec:gate-based-deployment}. A common parameterized form is:
\[
\begin{aligned}
H_{\mathrm{device}}(\boldsymbol{\lambda}, t)
  &= H_0 + \sum_k \lambda_k(t) H_k,
\end{aligned}
\]
where $H_0$ captures fixed device dynamics that are not directly controlled during execution; each $H_k$, for the device's supported control terms indexed by $k$, is either a coupling between physical subsystems (an interaction term) or a locally driven term (a control term); and $\lambda_k(t)$ is the time-dependent control coefficient that scales that term. The formula characterizes the device in three ways: the fixed term $H_0$, the set of terms $\{H_k\}$ the device can realize, and the admissible ranges and timing resolution of the control coefficients $\lambda_k(t)$. The latter two are the tunable parameters of the device, not specification-level quantities. Deployment feasibility therefore depends on whether the requested dynamics fall within these characterized ranges and on which physical subsystems can participate in the required terms \cite{peng2024simuq}.

\textbf{Physical subsystems and calibration.}
A \texttt{DeviceHamiltonian} acts on the \texttt{PhysicalQuantumSubsystem} instances owned by the processor; \model records this relation through \texttt{actsOn}. \texttt{PhysicalQuantumSubsystem} keeps the deployment view independent of qubit-only assumptions: a subsystem may be a qubit, qudit (a $d$-level quantum system with $d \ge 2$), bosonic mode (e.g., a photon or oscillator mode), fermionic mode (e.g., an occupied or unoccupied fermionic state), or another physical quantum unit. Its physical realization records the implementation technology, such as superconducting circuits, trapped ions, neutral atoms, photonic systems, or semiconductor spins. Similar to \texttt{ConnectivityTopology} in gate-based deployment, the processor's \texttt{InteractionTopology} gives a graph view of which physical subsystems can interact; its \texttt{SubsystemCoupling} instances generalize gate-based \texttt{QubitCoupling} to non-qubit subsystems. This topology complements \texttt{DeviceHamiltonian}: the Hamiltonian gives the mathematical form of each interaction term, while the topology records which of those terms the selected processor can physically realize.

As in gate-based deployment, \model keeps \texttt{CalibrationProfile} as a reference to the time-specific device characterization used for execution, rather than modeling the profile's internal fields. A deployment configuration selects one profile for execution, embedding, or feasibility assessment from the targeted backend's calibration history, tying each run to a definite device state. As discussed for simulator deployment, a calibration profile may inform a \texttt{NoiseModel}, but the two serve different roles: the calibration profile identifies device evidence, while the noise model specifies how a simulator approximates non-ideal behavior.

\textbf{Annealing Hamiltonians.}
A \texttt{QuantumAnnealer} specializes \texttt{AnalogQuantumProcessor} and uses physical qubits as its computational subsystems. With the device context established, annealing deployment turns to the computation: a logical problem Hamiltonian is mapped to physical qubits, combined with a driver Hamiltonian to form the requested evolution, and checked for realizability against the device Hamiltonian. Before deployment, an Ising energy model can be represented as the logical problem Hamiltonian
\[
H_{\mathrm{problem}}^{\mathrm{logical}}
= \sum_i h_i \sigma_i^z
  + \sum_{(i,j)} J_{ij} \sigma_i^z \sigma_j^z,
\]
where $i$ and $j$ identify logical variables represented in \model by \texttt{DecisionVariable} instances. The coefficients $h_i$ score individual variable assignments, while $J_{ij}$ scores pairwise compatibility or conflict between assignments \cite{kadowaki1998quantum,dwave2026annealing}. Together, these terms define the optimization landscape: assignments with lower energy represent better candidate solutions. Deployment must realize this logical Hamiltonian using the selected processor's physical qubits, supported couplings, and coefficient ranges, yielding the mapped Hamiltonian $H_{\mathrm{problem}}$. A typical forward-annealing evolution combines $H_{\mathrm{problem}}$ with a driver term:
\[
H_{\mathrm{anneal}}(s)
= A(s) H_{\mathrm{driver}} + B(s) H_{\mathrm{problem}},
\qquad s \in [0,1],
\]
where the parameter $s$ is the \texttt{ControlSchedule}'s progress parameter (Figure~\ref{fig:execution-evolution-uml}), not a \texttt{DecisionVariable}, and runs from $0$ at the beginning of the anneal to $1$ at its end. The functions $A(s)$ and $B(s)$ set the relative weights of the two terms: $A(s)$ weights the driver Hamiltonian, which helps the system explore candidate assignments, while $B(s)$ weights the problem Hamiltonian, whose low-energy states correspond to better solutions. A common driver applies a Pauli-$X$ operator to each logical variable, $H_{\mathrm{driver}} = -\sum_i \sigma_i^x$, where $i$ ranges over the same logical variables. This driver acts in the Pauli-$X$ basis, while $H_{\mathrm{problem}}$ is expressed in the Pauli-$Z$ basis: because Pauli-$X$ terms flip or mix candidate assignments they support exploration, whereas Pauli-$Z$ terms assign energies to those assignments so that lower-energy states represent better solutions.

As characterized above, $H_{\mathrm{device}}(\boldsymbol{\lambda}, t)$ describes what the selected processor can physically realize. Deployment checks whether $H_{\mathrm{driver}}$ and $H_{\mathrm{problem}}$ can each be expressed as linear combinations of the device's available operators $\{H_k\}$, and whether the schedule functions $A(s)$ and $B(s)$ can be realized by time-dependent settings of the control coefficients $\lambda_k(t)$. Constant factors and signs may differ between hardware and mathematical conventions.

\textbf{Annealing variable mapping and embedding.}
An \texttt{AnnealingDeploymentArtifact} specializes \texttt{AnalogDeploymentArtifact} and records how the logical problem's \texttt{DecisionVariable} instances are assigned to physical resources through one or more \texttt{VariableMapping} instances, each mapping one \texttt{DecisionVariable} to one or more \texttt{PhysicalQubit} instances. When the selected annealer's \texttt{InteractionTopology} supports every nonzero $J_{ij}$ interaction, each variable maps to a single physical qubit; otherwise it requires a connected set of qubits, commonly called minor embedding \cite{dwave2026annealing}. A valid mapping uses only physical qubits of the selected annealer, keeps the qubit sets of distinct variables disjoint, keeps each multi-qubit representation connected through \texttt{SubsystemCoupling} instances, and preserves each nonzero $J_{ij}$ through a supported coupling.

Beyond enabling structural feasibility checks before execution, these mappings are QSE-relevant because they affect resource estimates, execution reproducibility, and interpretation of sampled results. Mapping one \texttt{DecisionVariable} to multiple \texttt{PhysicalQubit} instances consumes additional physical qubits and introduces internal couplings. Coefficients $h_i$ and $J_{ij}$ that exceed the processor's supported range must be scaled down, which shrinks the energy differences between candidate solutions and makes the annealer more sensitive to noise. Consequently, embedding quality can affect whether a deployment fits the backend and how robust the sampled solutions are \cite{choi2008minor,dwave2026annealing}.

Overall, this view records what QSE needs to know to deploy analog and annealing computations. QSE deployment and reproducibility work treats backend identity, cloud execution, sample-count settings, compiler or SDK versions, calibration time, noise assumptions, and execution configuration as engineering context rather than incidental runtime detail \cite{weder2022qsdl,mauerer2022123reproduci,yue2026qready}. For annealing-oriented QSE, QUBO, Ising, and problem-Hamiltonian formulations already appear as software artifacts for test optimization, benchmarking, annealer execution, and program analysis \cite{zhang2026quantumoptimization,zhang2025empiricalstudies,wang2024testcasemini,yang2026isingtest,perezcastillo2023dynamic}. The analog-specific concepts in this view extend that deployment framing by making device dynamics, subsystem couplings, and variable-to-qubit mappings explicit.

\subsection{Application Examples}
\label{subsec:application-examples}
As a \textbf{representative embodied-AI case}, consider the hybrid quantum deep-reinforcement-learning TurtleBot navigation workflow studied by Hohenfeld et al.~\cite{hohenfeld2024quantumdeep}. The navigation application is an \texttt{Application} and hence a \texttt{SoftwareSystem}: classical \texttt{ClassicalComponent} instances encode sensor observations, maintain the navigation state, and execute the selected actions, while the simulated environment's physics engine detects collisions. Its \texttt{Requirement} instances include reaching the target while avoiding obstacles, together with limits on control-loop latency, training or execution cost, and acceptable failure probability. The navigation task is a \texttt{ComputationalProblem} with \texttt{ProblemKind::decision}, since the deliverable is an action choice. An \texttt{AlgorithmConfiguration} links that problem and its requirements to a \texttt{QuantumAlgorithm}: a hybrid reinforcement-learning method in which a parameterized quantum circuit (PQC) approximates the action-value (Q) function. The configuration is implemented by a \texttt{QuantumProgram}, a \texttt{SoftwareComponent}, that encodes each observation, binds its values to circuit parameters, and constructs a gate-based \qcs represented by a \texttt{GateBasedComputation} containing a \texttt{QuantumCircuit}. The circuit applies operations to \texttt{LogicalQubit} instances and measures the resulting state to provide action values to the classical learning loop, which updates the PQC parameters.

For execution and deployment, a \texttt{DeploymentConfiguration} selects a \texttt{GateBasedDeploymentArtifact}, a \texttt{QuantumBackend} of type \texttt{SimulatorBackend}, and a \texttt{SimulationConfiguration}; an optional \texttt{NoiseModel} records non-ideal simulation assumptions. TensorFlow Quantum with Cirq/Qsim provides the quantum-side simulator, while PyBullet supplies the external embodied-navigation environment. At each step of an evaluation episode, the current observation induces a \texttt{QuantumExecution} with a recorded \texttt{sampleCount}; the execution instantiates a \texttt{QuantumSystem} and returns an \texttt{ExecutionResult} containing \texttt{MeasurementRecord} instances. A classical component decodes these records into a motion command and sends it to the simulated robot, which executes it and detects collisions. The resulting state transition supplies the next input and repeats the loop. This mapping keeps the reusable program, gate-level specification, simulator deployment, runtime evidence, and classical control boundary distinct, and it makes clear that the case provides simulator-based embodied evaluation rather than physical-QPU evidence.

Consider also a \textbf{hybrid test-suite optimization application} which addresses the SE problem of selecting a small test-case subset while preserving coverage. Its requirements specify the desired selection behavior and constrain quantum-specific feasibility and quality attributes, including cost, available qubits, circuit depth, backend access, execution budget, and acceptable failure probability. The application owns a \texttt{ComputationalProblem} with \texttt{ProblemKind::optimization}. These requirements constrain an \texttt{AlgorithmConfiguration} that selects QAOA, a \texttt{QuantumAlgorithm} using \texttt{variational} and \texttt{HamiltonianEvolution} techniques. The configuration is implemented by a \texttt{QuantumProgram}, a \texttt{SoftwareComponent} of the application, alongside a \texttt{ClassicalComponent} that prepares inputs and checks coverage.

The \texttt{QuantumProgram} constructs a \qcs realized as a \texttt{GateBasedComputation} containing a \texttt{QuantumCircuit}. Its \texttt{StatePreparation}, \texttt{Gate}, and \texttt{Measurement} operations act on \texttt{LogicalQubit} instances and realize the QAOA cost and mixer Hamiltonians referenced via \texttt{computationHamiltonians}. A \texttt{DeploymentConfiguration} selects a \texttt{GateBasedDeploymentArtifact} with \texttt{QubitMapping} instances, a \texttt{QuantumBackend}, calibration, and noise context. A \texttt{QuantumExecution} runs the artifact with a \texttt{sampleCount} and returns an \texttt{ExecutionResult} containing \texttt{MeasurementRecord} instances; the classical component interprets these records as candidate subsets and checks them against the requirements. The same problem may use a \texttt{QuantumAnnealingComputation} over an \texttt{EnergyModel}, showing \texttt{ProblemKind} does not determine the paradigm \cite{wang2024quantumappro,wang2024testcasemini}.

\section{Assessment of \model for QSE}
\label{sec:literature-assessment}
This section assesses whether \model provides the QC foundation the QSE activities require. It does not ask whether \model replaces those activities.
We organize the literature into three groups by their role in the assessment, rather than by an exhaustive taxonomy of QSE research. 
1) \textbf{High-level summaries and reviews}: QSE surveys, roadmaps, and lifecycle studies, e.g., \cite{piattini2022landscape,weder2022qsdl,dwivedi2024qsesurvey,murillo2025qseroadmap}, identify the relevant activity space to establish which SE tasks the assessment must cover.
2) \textbf{Overall organization and structure}: requirements, architecture, and modeling studies, e.g., \cite{yue2023towardsquant,khan2023qsa,perezdelgado2022quml,gemeinhardt2021towardsmodel,guo2025quanumltowar,guo2025m2qcodeamode}, expose the SE--QC bridge at the level of artifacts and decisions, and show how application needs, system structures, and development models refer to QC concepts. 
3) \textbf{Fine-grained QSE phases and goals}: studies of testing, debugging, runtime behavior, reproducibility, and maintenance, e.g., \cite{garcia2023qst,fortunato2024vv,oldfield2025qst,paltenghi2023morphqmetamo,wang2021qdiffdiffere,mauerer2022123reproduci}, provide operational evidence that helps identify the QC concepts QSE activities must expose when artifacts are executed, evaluated, and evolved.

Together, these groups define an assessment chain from the QSE task scope, through the SE--QC bridge, to the runtime and maintenance evidence that the foundation should support.
The unit of analysis is therefore one QSE task and its required SE--QC bridge.
Accordingly, QC concepts required by a task should be represented in \model, while task-specific artifacts and procedures, such as test cases, test oracles, defect reports, change requests, evaluation metrics, and process activities, remain in a separate QSE activity layer that uses \model as its QC vocabulary.
Symmetrically, when a QC concept already represented in \model is only weakly addressed by current QSE studies, we interpret this as an underdeveloped part of the activity layer rather than evidence that the conceptual foundation is incomplete.

The following subsections apply this assessment logic to the QSE activities in which SE artifacts and decisions should be connected to QC-specific assumptions, computations, execution evidence, and deployment context.

\subsection{Requirements}
\label{subsec:assessment-requirements}

Requirements engineering elicits, analyzes, specifies, validates, and manages stakeholder needs and constraints for an intended system. In QSE, it is essential because the suitability of a computational problem for QC, the allocation between classical and quantum computation, and feasibility under hardware constraints should be understood before algorithm, circuit, and backend decisions are made.

In the literature, Saraiva et al.~\cite{saraiva2021nonfunctional} argued the importance of hardware-related non-functional requirements, including limits on qubit count, circuit depth, qubit connectivity, and ancilla-qubit overhead. Yue et al.~\cite{yue2023towardsquant} distinguished quantum, classical, and hybrid requirements, connected application needs to quantum-algorithm selection, and identified gate and ancilla-qubit estimates, qubit and circuit-depth limits, and hardware-error-related reliability concerns. Stein et al.~\cite{stein2024requirementsmanufacturing} additionally noted that an application may involve heterogeneous computing resources, e.g., CPU, GPU, and QPU.
In \model, all these aspects are represented by \texttt{SoftwareSystem}, \texttt{Application}, \texttt{Requirement}, \texttt{QuantumProgram}, \texttt{QuantumAlgorithm}, \texttt{QuantumCircuit}, \texttt{Qubit}, \texttt{QubitPurpose}, and \texttt{QPUBackend}.

However, we observe that the quantum requirements engineering literature usually starts from gate-based programs and circuits. It recognizes hardware constraints but does not systematically represent the \texttt{ComputationalProblem}, the composition and development of a quantum application, or the choice among computation paradigms. Our position is that requirements engineering should first establish the computational problem, applicable quantum algorithms or techniques, candidate \qcs{} paradigms, and relevant backend characteristics, before committing to a realization. For gate-based computation, this means going beyond circuit depth, width (qubit count), and gate counts: requirements should distinguish \texttt{LogicalQubit} demand from \texttt{PhysicalQubit} availability and constrain connectivity, calibration profiles, noise models, and backend mappings. \model supplies these concepts.

\subsection{Architecture Design}
\label{subsec:assessment-architecture-design}
Architecture design structures a software system and selects technologies through which it realizes its requirements. Modeling notations, tools, and methodologies support engineers in representing, analyzing, communicating, and making these decisions for quantum applications.

More than a decade ago, Wecker and Svore~\cite{wecker2014liqui} recognized the need to raise the abstraction level above quantum physics and linear algebra when designing quantum algorithms. $LIQUi|>$ provides this support through a domain-specific language, while its abstractions remain centered on qubits, gates, circuits, and operations for programming, compilation, and simulation. Notably, the work explicitly addresses both gate-based and adiabatic QC; its implementation nevertheless focuses on gate-based QC.

The work by Ali and Yue~\cite{ali2020modelingquan} is an early effort to understand quantum programs from an SE perspective. The authors proposed to use the UML state machine notation as a high-level description of program execution. One noticeable limitation concerns the very restricted definition of \texttt{State} itself, which omits the measurement basis and represents quantum states primarily through computational-basis outcome probabilities.
QuanUML \cite{guo2025quanumltowar} extends UML class and sequence diagram notations with QC concepts, including qubits, gates, and inter-qubit messages. Its aim is to specify quantum algorithms at low level and generate quantum programs automatically \cite{guo2025m2qcodeamode}. However, on the notation side, representing gates as messages exchanged between qubits can be misleading. For instance, a CNOT gate is one joint operation with control and target operands, not two messages sent between the participating qubits.
Q-UML \cite{perezdelgado2022quml} is another UML-based notation for specifying quantum algorithms. It extends UML graphical notations to distinguish quantum from classical elements, aiming to facilitate the understanding of an algorithm's high-level logic and classical--quantum workflow. However, some Q-UML constructs can suggest that quantum states and operations are ordinary values or messages that can be directly set, retrieved, or exchanged. On actual quantum hardware, programs manipulate qubits through quantum operations and obtain classical information through measurement, and arbitrary internal states are not directly readable.

In the literature, gate-based QC, typically expressed as quantum circuits, is usually taken as the default. Current modeling notations remain limited in scope and largely low-level: they primarily describe quantum-algorithm and circuit elements, with little support for quantum-application design, computation-paradigm selection, or deployment-architecture design. Raising the level of abstraction, particularly through UML-based notations, is valuable; however, these notations must preserve QC semantics rather than recast quantum states, operations, or measurements as ordinary classical objects or messages.

\model retains the literature's core QC concepts---software systems and components, quantum programs and algorithms, circuits, operations, qubits, mappings, compiled circuits, and backends---but adds end-to-end design reasoning. It links \texttt{Application}, \texttt{SoftwareSystem}, \texttt{Requirement}, and \texttt{ComputationalProblem} to a selected \texttt{QuantumAlgorithm} through an \texttt{AlgorithmConfiguration}, which is realized by a paradigm-specific \qcs. This \qcs may be a \texttt{GateBasedComputation} with a \texttt{QuantumCircuit}, a \texttt{QuantumAnnealingComputation} with an \texttt{EnergyModel}, an \texttt{AnalogComputation} with a \texttt{HamiltonianModel} and \texttt{ControlSchedule}, or a \texttt{MeasurementBasedComputation} with a \texttt{MeasurementPattern}, subject to problem, requirements, backend, and deployment constraints. \model therefore supplies the QC concepts that architecture-design artifacts can reference, including the application, paradigm, and deployment concerns largely absent from the reviewed literature. It intentionally leaves generic architectural constructs---components, interfaces, connectors, and decisions---to UML, ADLs, and established SE methods, so that concrete notations can combine these constructs with \model's QC concepts.

\subsection{Code Generation}
\label{subsec:assessment-code-generation}
Quantum program generation realizes the decisions made during requirements and architecture design as executable quantum and hybrid software. This subsection considers the transformations from requirements and design models to quantum programs, paradigm-specific specifications, and backend-adapted artifacts, and the QC concepts that these transformations must preserve.

M2QCode~\cite{guo2025m2qcodeamode} generates quantum programs from low-level UML models, while Jim\'enez-Navajas et al.~\cite{jimenez2025umlcodegeneration} proposed to generate Python-Qiskit artifacts for hybrid quantum-classical software systems modeled in Q-UML~\cite{perezdelgado2022quml}.
Gemeinhardt et al.~\cite{gemeinhardt2024composition} supported compositional circuit design and automated code generation through composite quantum gates for circuit reuse and a loop operation for control flow.
All three approaches target gate-based circuits and use abstractions close to programming, such as qubits, gates, circuits, and control-flow constructs. They therefore support transformations from low-level design models to quantum programs, but provide limited support for higher-level application design or reasoning about alternative computation paradigms.
QuantME~\cite{weder2020quantumworkflows} generates executable hybrid workflow artifacts from quantum-enhanced process models. Unlike circuit-code generators, it focuses on orchestrating and deploying a quantum application, which we discuss in Section~\ref{subsec:assessment-execution-deployment}.
Emerging LLM-based work explores generation at different levels. Siavash and Moin~\cite{siavash2025llmcodegeneration} proposed to generate Qiskit code from UML models using retrieval-augmented generation, whereas QAgent~\cite{fu2025qagent} autonomously generates OpenQASM programs. QPipe~\cite{tao2026qpipe} is an LLM-based multi-agent system that operates at the application level by translating natural-language optimization requirements into executable quantum applications with traceable artifacts for requirement parsing, formulation, code generation, review, execution, and verification.

\model covers most QC concepts used in the above literature, including quantum programs, circuits, operations, qubits, registers, parameters, and compiled circuits. Gemeinhardt et al.'s composite gates and loop operation~\cite{gemeinhardt2024composition} suggest two concepts that \model does not adopt. \model does not introduce a separate \texttt{CompositeQuantumOperation}, because operation composition is represented at the circuit level through ordered \texttt{OperationApplication} instances that invoke \texttt{QuantumOperation} definitions; nor does it model loops, which are implementation-level control-flow constructs expressible in a programming or behavioral notation rather than as QC concepts. Moreover, \model covers multiple QC paradigms rather than only gate-based circuits and represents quantum systems, states, deployment, execution, and evolution. For code generation, \texttt{DeploymentConfiguration} provides target-specific information for compilation and backend selection, and \texttt{QuantumState}, \texttt{StatePreparation}, \texttt{Gate}, and \texttt{Measurement} guide the generation and validation of state preparation, transformation, and measurement instructions.

\subsection{Execution and Deployment}
\label{subsec:assessment-execution-deployment}

Execution and deployment make generated quantum and hybrid artifacts operational by binding them to executable workflows, service interfaces, backend-specific deployment artifacts, and runtime evidence.

QPipe~\cite{tao2026qpipe}, discussed above, also executes and verifies the applications it generates. QuantME~\cite{weder2020quantumworkflows} specifies hybrid quantum circuit-based workflows and transforms them into executable workflow artifacts. Its workflows refer to QC concepts such as \texttt{QuantumCircuit}, \texttt{NoiseModel}, \texttt{DeploymentConfiguration}, \texttt{QPUBackend}, \texttt{QuantumExecution}, and \texttt{ExecutionResult}, while generic workflow constructs remain the responsibility of established workflow and SE notations. Follow-up work further distinguishes application-workflow orchestration from deployment orchestration \cite{weder2021orchestrations}; QC patterns can also automate the construction and adaptation of hybrid workflows \cite{beisel2025patternbased}.

As already discussed in Sections~\ref{subsec:assessment-requirements} and~\ref{subsec:assessment-architecture-design}, some execution and deployment decisions need to be made in earlier QSE activities. However, the literature remains largely circuit-based, with other QC paradigms insufficiently covered. \model makes an initial effort to conceptually connect \texttt{Requirement} and \texttt{Application} to an \texttt{AlgorithmConfiguration} and paradigm-specific \qcs, and then to \texttt{DeploymentConfiguration}, \texttt{QuantumBackend}, \texttt{QuantumExecution}, and \texttt{ExecutionResult} across paradigms.
These links support feedback across QSE activities. Generated code, compilation, workflow adaptation, or backend assessment may reveal constraints that require revised requirements or design decisions \cite{weder2022qsdl,yue2026qready,murillo2025qseroadmap}. Furthermore, as shown in Figure~\ref{fig:execution-evolution-uml}, \texttt{ExecutionResult} records raw evidence, while \texttt{ErrorMitigationApplication} applies an \texttt{ErrorMitigationStrategy} and produces a separate \texttt{MitigatedResult}. Comparing the two helps assess noise effects and mitigation effectiveness \cite{muqeet2024mitigatingno}. \model also records the QC-specific \texttt{sampleCount} of a \texttt{QuantumExecution}; broader study-level measures, such as independent experimental repetitions, remain outside its scope \cite{zhang2025empiricalstudies}.

\subsection{Quality Assurance}
\label{subsec:assessment-quality-assurance}

Quality assurance activities, including testing, verification, and debugging, are much more mature than other SE activities in the QSE literature. Quantum software testing (QST) is the field that provides the richest body of empirical evidence such as mutation testing (Muskit)~\cite{mendiluze2021muskitamutat}, metamorphic testing (MorphQ)~\cite{paltenghi2023morphqmetamo}, combinatorial testing (QuCAT)~\cite{wang2023qucatacombin}, differential testing (QDiff)~\cite{wang2021qdiffdiffere}, statistical and projective-measurement testing~\cite{sato2025buglocating,oldfield2025qst}, runtime assertion checking (ProQ and Rovara et al.)~\cite{li2020proq,rovara2025runtimeassertions}, static analysis (LintQ)~\cite{paltenghi2024analyzingqua}, and bug benchmarks (Bugs4Q)~\cite{zhao2023bugs4qabench}. Recent methodological analysis further reports that empirical QST studies vary substantially in their programs under test, inputs, oracles, baselines, shot configurations, execution repetitions, backend choices, and artifact support \cite{li2026qstmethodology}.

Despite this maturity, the QST literature is predominantly gate-based and circuit-centered. This emphasis appears in both the program representations and platforms used by existing techniques, with substantial reliance on Qiskit and gate-model programs \cite{garcia2023qst,fortunato2024vv}. Consequently, the reviewed evidence is strongest for gate-based programs, while testing, verification, and debugging for analog, annealing, and measurement-based paradigms remain less developed. The literature also does not yet provide a general testing solution for hybrid quantum applications or sufficiently bridge the interpretation of results obtained from simulator and QPU backends. These gaps identify areas that \model can help address: \model relates these testing concerns to QC concepts across its operation, qubit, state, execution and evolution, and deployment views. A gate-based program or circuit under test can refer to \texttt{QuantumProgram}, \texttt{QuantumCircuit}, \texttt{OperationApplication}, \texttt{ParameterBinding}, \texttt{Measurement}, \texttt{Reset}, and \texttt{OperationContract}. Test inputs and expected states or properties can refer to \texttt{LogicalQubit}, \texttt{QuantumState}, \texttt{QuantumSystemState}, and \texttt{SystemBasis}. Test execution and evidence can refer to \texttt{QuantumExecution}, \texttt{sampleCount}, \texttt{ExecutionResult}, \texttt{MeasurementRecord}, \texttt{MeasurementOutcome}, \texttt{NoiseModel}, \texttt{QuantumBackend}, \texttt{CalibrationProfile}, and \texttt{DeploymentConfiguration}. These concepts allow \model to express test oracles for non-gate paradigms in terms of intended evolution, observables, and outcome distributions, complementing gate-sequence oracles. More generally, linking the specification under test, the computation being executed, the measurement evidence, and the backend or noise assumptions makes QST results interpretable and supports diagnosis.

Li et al.'s ProQ~\cite{li2020proq} expresses quantum-state properties as predicates over selected quantum variables and checks them through projective measurements, with transformations that make the predicates executable under measurement restrictions. Rovara et al.~\cite{rovara2025runtimeassertions} define equality and superposition assertions and propose an automated framework that translates them into measurements, splits programs into executable slices for multiple assertions, reduces evaluation overhead, and evaluates the outcomes under noise on real quantum computers. More generally, \model represents assertion specifications as properties over an expected \texttt{QuantumState} or \texttt{QuantumSystemState} associated with selected \texttt{LogicalQubit}s or a \texttt{QuantumRegister}. Assertions may be associated with selected points in a \texttt{QuantumEvolution} or \texttt{QuantumExecution}, with evidence represented by simulator-specific \texttt{StateSnapshot}s or by hardware \texttt{ExecutionResult}s and measurement records. These assertion works are again gate-based. Future runtime assertions for analog quantum computations can build on \model's \texttt{QuantumEvolution}, \texttt{HamiltonianModel}, \texttt{DeviceHamiltonian}, state, and execution concepts to constrain observables, energy values, state overlap, or outcome distributions during or after Hamiltonian evolution.

Debugging work locates and explains faults after observing anomalous behavior. Sato et al.~\cite{sato2025buglocating} divided a quantum program into segments and used cost-based search, early determination, finalization, and look-back strategies to localize faults under the cost and statistical constraints of repeated quantum executions. QDiff~\cite{wang2021qdiffdiffere} generates semantically equivalent circuit variants, filters them by circuit and hardware characteristics, and compares their outcome distributions on simulators and QPUs to expose software-stack instabilities. Bugs4Q~\cite{zhao2023bugs4qabench} provides manually validated and reproducible Qiskit bugs with buggy/fixed versions and reproducing tests for controlled debugging studies. The QC concepts used by these works, including \qcs, \texttt{QuantumCircuit}, \texttt{OperationApplication}, \texttt{CompiledCircuit}, \texttt{QuantumExecution}, \texttt{sampleCount}, \texttt{QuantumBackend}, \texttt{ExecutionResult}, and \texttt{MeasurementOutcome}, are covered by \model. \model additionally supports non-gate debugging through \texttt{HamiltonianModel}, \texttt{ControlSchedule}, \texttt{QuantumEvolution}, and observable-based evidence, and represents raw versus mitigated results through \texttt{ErrorMitigationApplication}, \texttt{ErrorMitigationStrategy}, and \texttt{MitigatedResult}. The debugging literature covers search procedures, diagnoses, and bug benchmarks, which remain SE debugging artifacts rather than core QC concepts, and are hence not covered by \model.

\section{Conclusion}
\label{sec:conclusion}
Quantum software engineering (QSE) needs a precise account of quantum computing concepts that supports software engineering decisions and evidence. We introduced \model, a conceptual model that distinguishes applications, algorithms, programs, computation specifications, executions, and deployment artifacts.
Our assessment shows that current QSE work is largely gate-based, circuit-centered, and close to programming-level abstractions, with limited support for application design, alternative paradigms, hybrid applications, deployment reasoning, and interpreting simulator and hardware results. \model covers the QC concepts these works use and extends them with explicit representations for alternative paradigms, quantum evolution, deployment context, execution evidence, and raw versus mitigated results. It therefore provides the QC foundation QSE activities can reference without replacing established SE methods, and remains incomplete by design, to evolve with QC technologies and QSE practices.


\section*{Acknowledgment}
This work is supported by the fundamental research funds for the central universities.

\bibliographystyle{IEEEtran}
\bibliography{references}

\end{document}